\input{aipcheck}
\documentclass[
    ,final            
  ]
  {aipproc}

\layoutstyle{8x11single}

\usepackage{graphicx,xcolor}

\newcommand{\beq}{\begin{equation}}
\newcommand{\eeq}{\end{equation}}
\newcommand{\bea}{\begin{eqnarray}}
\newcommand{\eea}{\end{eqnarray}}

\begin{document}

\title{The $\gamma^* \gamma^*$ total cross section in NLA BFKL}

\classification{12.38.Bx,12.38.Cy,13.85.Lg}
\keywords      {Photon-photon total cross section, BFKL resummation}
\author{Dmitry Yu.~Ivanov}{address={Sobolev Institute of Mathematics 
and Novosibirsk State University, 630090 Novosibirsk}}

\author{Beatrice Murdaca}{address={Dipartimento di Fisica, Universit\`a della 
Calabria, and Istituto Nazionale di Fisica Nucleare, Gruppo collegato di
Cosenza, Arcavacata di Rende, I-87036 Cosenza, Italy}
}

\author{Alessandro Papa}{address={Dipartimento di Fisica, Universit\`a della 
Calabria, and Istituto Nazionale di Fisica Nucleare, Gruppo collegato di
Cosenza, Arcavacata di Rende, I-87036 Cosenza, Italy}
}

\begin{abstract}
We study the $\gamma^* \gamma^*$ total cross section
in the NLA BFKL approach. We have extracted the NLO corrections to
the photon impact factor from two recent papers of Balitsky and Chirilli and 
Chirilli and Kovchegov and used them to build several representations 
of the total cross section, equivalent within the NLA.
We have combined these different representations with two among the most 
common methods for the optimization of a perturbative series, namely PMS and 
BLM, and compared their behavior with the energy with the only available 
experimental data, those from the LEP2 collider.
\end{abstract}

\maketitle


\section{Introduction}

\begin{figure}[t]
\begin{minipage}{0.50\textwidth}
\centering
\includegraphics[scale=0.7]{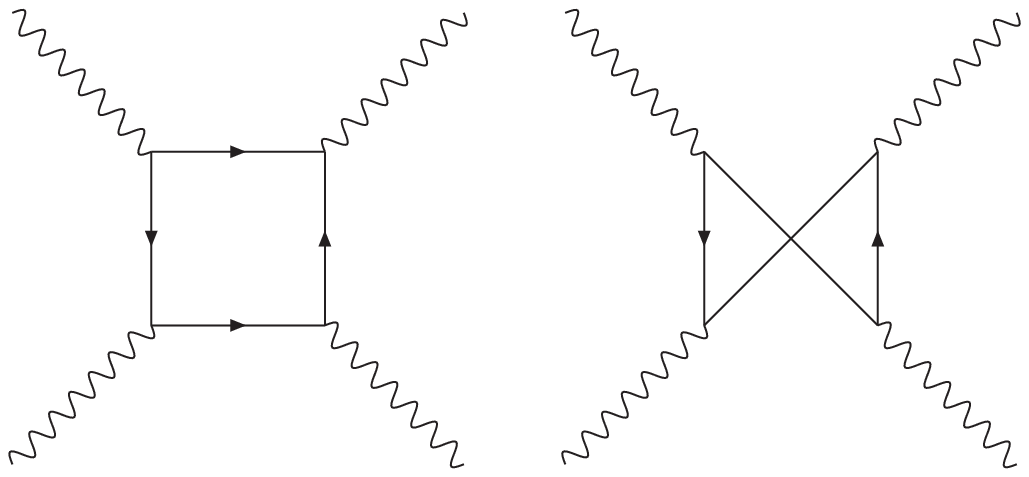}
\end{minipage}
\begin{minipage}{0.50\textwidth}
\centering
\includegraphics[scale=0.65]{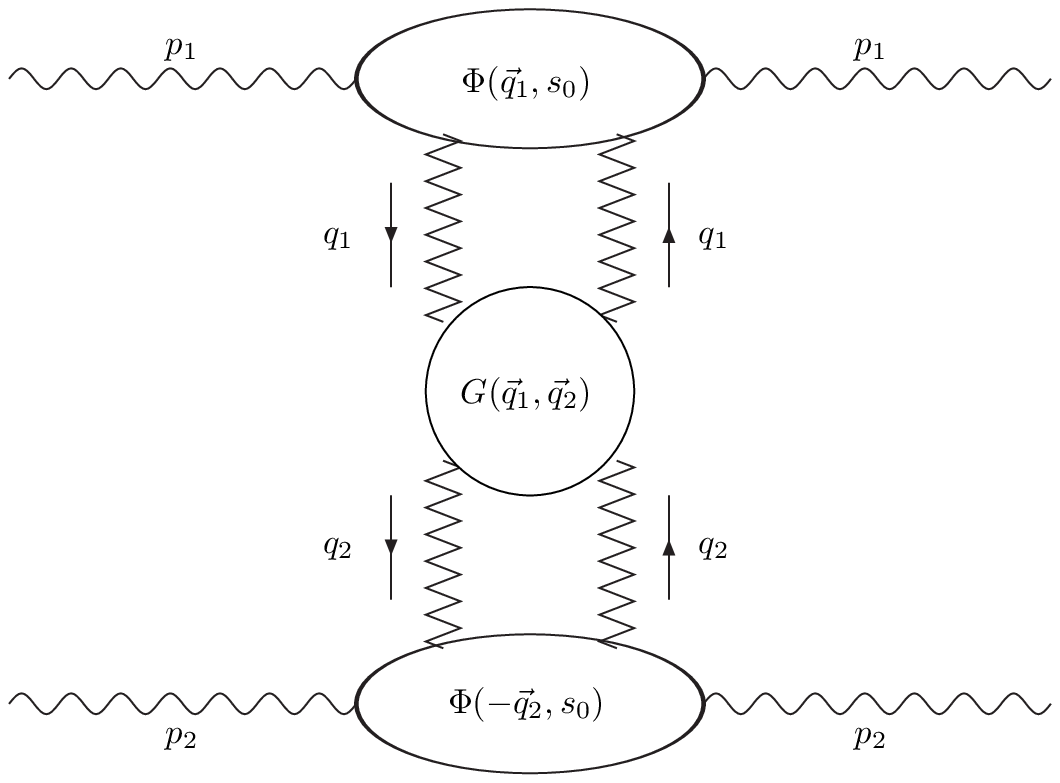}
\end{minipage}
\caption{(Left) Quark box LO diagrams. 
(Right) Schematic representation of the $\gamma^*(p_1)\, \gamma^*(p_2)$ 
forward scattering.}
\label{fig:diagrams}
\end{figure}

The total cross section for the collision of two off-shell photons with large 
virtualities is an important test ground for perturbative QCD.
At a fixed order of $\alpha_s$ and at low energies, the dominant contribution
comes from the quark box, calculated at the leading-order (LO)
in Refs.~\cite{Budnev-Schienbein} (see Fig.~\ref{fig:diagrams} 
(left)) and at the next-to-LO (NLO) in Ref.~\cite{Cacciari:2000cb}.
In Ref.~\cite{BL03} the resummation of double logs appearing in the NLO
corrections to the quark box was also studied.
At higher energies, the gluon exchange in the $t$-channel becomes dominant
and gives terms with powers of energy logs which must be resummed to all orders.

The procedure for this resummation in the leading logarithmic approximation 
(LLA) (terms $(\alpha_s\ln(s))^n$) and in the next-to-leading 
logarithmic approximation (NLA) (terms $\alpha_s(\alpha_s\ln(s))^n$)
has been established within the BFKL approach~\cite{BFKL}: the imaginary part 
of an amplitude (and, hence, a total cross section) for a large-$s$ hard 
collision process reads as the convolution of the Green's function of two 
interacting Reggeized gluons with the impact factors (IFs) of the colliding 
particles (see Fig.~\ref{fig:diagrams} (right)).

The Green's function is universal and is known in the NLA for singlet color
representation in the $t$-channel and forward scattering~\cite{NLA-kernel}.
The leading order (LO) photon IF is known since long, but it took 
years to calculate the next-to-LO (NLO) one~\cite{gammaIF}. Its lengthy 
expression, in the momentum representation, was published over a few years in 
pieces, some of them available only in the form of a numerical code, thus 
making it of limited practical use. Indeed, so far, the inclusion 
of BFKL resummation effects in the NLA calculation of the $\gamma^* \gamma^*$ 
total cross section was carried out approximately, taking NLA Green's function 
and LO IFs~\cite{Brodsky,Caporale2008,Zheng}.
A few months ago, the NLO photon IF was calculated in the coordinate 
space and then transformed to the momentum representation and to the Mellin (or 
$\gamma$-representation)~\cite{Balitsky2012} (see also~\cite{Chirilli2014}). 
It turns out that its expression is very simple in all representations, thus 
confirming (see, for instance,~\cite{coordinate}) that the use of the coordinate
representation leads to simple expressions, which, in the momentum 
representation, would result after not so obvious cancellations.

Now all ingredients are available to build the $\gamma^* \gamma^*$ total
cross in NLA BFKL.
Previous studies based on the NLA BFKL approach, such as the photoproduction of 
two light vector mesons~\cite{IKP04,mesons_1-2,mesons_3} and the production of 
Mueller-Navelet jets~\cite{MN_IF,MN_sigma}, have clearly shown that NLA 
expressions for an observable (such as a cross-section or an azimuthal 
correlation), though being formally equivalent up to subleading terms, may lead 
to somewhat different numerical estimates. This derives from the fact that
NLO BFKL corrections are typically of opposite sign with respect to the LO and 
large in absolute value and calls for (i) an optimization procedure for the 
perturbative series and (ii) a check of the stability of the numerical result 
under change of the representation, within a large enough class of 
NLA-equivalent expressions.

In this paper we compare several NLA-equivalent representations of 
the $\gamma^* \gamma^*$ total cross section, using two methods of optimization 
of the perturbative series, the principle of minimal 
sensitivity (PMS)~\cite{Stevenson} and the Brodsky-Lepage-Mackenzie (BLM) 
method~\cite{BLM}. Results will be contrasted with the experimental data 
obtained at LEP2~\cite{Achard:2001kr,Abbiendi:2001tv}.

\section{BFKL contribution to the $\gamma^* \gamma^*$ total cross section}

The total cross section of two unpolarized photons with virtualities
$Q_1$ and $Q_2$ in LLA BFKL and in the Mellin-representation (also
said $\gamma$- or $\nu$-representation), it is given by (see, for instance, 
Ref.~\cite{Brodsky}):
\begin{equation}
\sigma^{\gamma^{*} \gamma^{*}}_{\rm tot} (s,Q_1,Q_2) =
\sum_{i,k=T,L} \frac{1}{(2 \pi)^2 Q_1 Q_2}
\int\limits^{+\infty}_{-\infty} d\nu \left(\frac{Q_1^2}{Q_2^2}\right)^{i\nu}
F_i(\nu) F_k(-\nu) \left(\frac{s}{s_0}\right)^{\bar \alpha_s \chi(\nu)} \; ,
\label{sigmaLO}
\end{equation}
where $\bar \alpha_s\equiv \alpha_s(\mu_R) N_c/\pi$, with $N_c$ the number
of colors, $\chi(\nu)=2\psi(1)-\psi\left(1/2+i\nu\right)-\psi\left(1/2
-i\nu\right)$ and
\[
F_T(\nu) = F_T(-\nu)= \alpha \alpha_s \left( \sum_q e_q^2 \right)
\frac{\pi}{2} \frac{(\frac{3}{2} - i\nu)(\frac{3}{2}+ i\nu)
\Gamma^2(\frac{1}{2} - i\nu)\Gamma^2(\frac{1}{2}+i\nu)}
{\Gamma(2-i\nu) \Gamma(2+i\nu)}
= \alpha \alpha_s \left( \sum_q e_q^2 \right)
\frac{\pi^2}{8}\frac{9+4\nu^2}{\nu\left(1+\nu^2\right)}
\frac{\sinh \left(\pi\nu\right)}{\cosh^2(\pi\nu)},
\]
\[
F_L(\nu) = F_L(-\nu)= \alpha \alpha_s \left( \sum_q e_q^2 \right) \pi
\frac{\Gamma(\frac{3}{2} - i\nu)\Gamma(\frac{3}{2} + i\nu)
\Gamma(\frac{1}{2} - i\nu) \Gamma(\frac{1}{2} + i\nu)}
{\Gamma(2-i\nu) \Gamma(2+i\nu)} 
= \alpha \alpha_s \left( \sum_q e_q^2 \right)
\frac{\pi^2}{4}\frac{1+4\nu^2}{\nu\left(1+\nu^2\right)}
\frac{\sinh \left(\pi\nu\right)}{\cosh^2(\pi\nu)}
\]
are the LO IFs for transverse and longitudinal polarizations,
respectively. Here, $\alpha$ is the electromagnetic coupling constant, the 
summation extends over all active quarks (taken massless) and $e_q$ is the 
quark electric charge in units of the electron charge. In the LLA BFKL cross 
section~(\ref{sigmaLO}) the scales $\mu_R$ and $s_0$ are not fixed.

Following Refs.~\cite{mesons_1-2}, it is possible to write down the NLA BFKL 
cross section as follows:
\[
\sigma^{\gamma^{*} \gamma^{*}}_{\rm tot} (s,Q_1,Q_2,s_0,\mu_R)
= \frac{1}{(2 \pi)^2 Q_1 Q_2}
\int\limits^{+\infty}_{-\infty} \!d\nu\! \left(\frac{Q_1^2}{Q_2^2}\right)^{i\nu}\!\!
\left(\frac{s}{s_0}\right)^{\bar \alpha_s(\mu_R) \chi(\nu)}
\!\!\!\!\!\sum_{i,k=T,L}F_i(\nu)F_k(-\nu) \left\{1 +\bar\alpha_s(\mu_R)
\left(\frac{F_i^{(1)}(\nu,s_0,\mu_R)}{F_i(\nu)} \right.\right.
\]
\beq
\left. +\frac{F_k^{(1)}(-\nu,s_0,\mu_R)} {F_k(-\nu)}\right) 
\left. + \bar \alpha_s^2(\mu_R) \ln\left(\frac{s}{s_0}\right) \left[ \bar
\chi(\nu)+\frac{\beta_0}{8N_c}\chi(\nu)\left(-\chi(\nu)+\frac{10}{3}
+ 2\ln\frac{\mu_R^2}{Q_1Q_2} \right) \right]\right\} \; ,
\label{sigmaNLO}
\eeq
where
\[
\bar\chi(\nu)=-\frac{1}{4}\left[\frac{\pi^2-4}{3}\chi(\nu)-6\zeta(3)-
\chi^{\prime\prime}(\nu)-\frac{\pi^3}{\cosh(\pi\nu)}
+ \frac{\pi^2\sinh(\pi\nu)}{2\,\nu\, \cosh^2(\pi\nu)}
\left(3+\left(1+\frac{n_f}{N_c^3}\right)\frac{11+12\nu^2}{16(1+\nu^2)}
\right) +\,4\,\phi(\nu)
\right] \, ,
\]
\[
\phi(\nu)\,=\,2\int\limits_0^1dx\,\frac{\cos(\nu\ln(x))}{(1+x)\sqrt{x}}
\left[\frac{\pi^2}{6}-\mbox{Li}_2(x)\right]\, , \;\;\;\;\;
\mbox{Li}_2(x)=-\int\limits_0^xdt\,\frac{\ln(1-t)}{t} \, ,
\]
$n_f$ is the number of active quarks, $F^{(1)}_{L,T}(\nu,s_0,\mu_R)$ are the
NLO corrections to the longitudinal/transverse photon IF in the
$\nu$-representation and $\beta_0= 11 N_c/3-2n_f/3$.

By comparing Eq.~(\ref{sigmaNLO}) with the $\gamma^*\gamma^*$ cross section 
obtained in the Wilson-line operator expansion scheme by Chirilli and 
Kovchegov~\cite{Chirilli2014}, we can extract the NLO parts of the photon 
IFs (for details, see~\cite{IMP14}),
\[
\frac{F_T^{(1)}(\nu,s_0,\mu_R)}{F_T(\nu)}=
\frac{\chi(\nu)}{2}\ln\frac{s_0}{Q^2}+\frac{\beta_0}{4 N_c}\ln\frac{\mu_R^2}{Q^2}
+\frac{3 C_F}{4 N_c}-\frac{5}{18}\frac{n_f}{N_c}+\frac{\pi^2}{4}+\frac{85}{36}
-\frac{\pi^2}{\cosh^2(\pi\nu)}-\frac{4}{1+4\nu^2}
+\frac{6\chi\left(\nu\right)}{9+4\nu^2}
\]
\beq
+\frac{1}{2\left(1-2 i\nu\right)}
-\frac{1}{2\left( 1+2 i\nu\right)}
-\frac{7}{18\left( 3+2 i\nu\right)}+\frac{20}{3\left(3+2 i\nu\right)^2}
-\frac{25}{18\left(3-2 i\nu\right)}
\label{ifT}
\eeq
\[
+\frac{1}{2}\chi\left(\nu\right)\left[\psi\left(\frac{1}{2}-i\nu\right)
+2\psi\left(\frac{3}{2}-i\nu\right)-2\psi\left(3-2 i\nu\right)
-\psi\left(\frac{5}{2}+i\nu\right)\right]
\]
and
\[
\frac{F_L^{(1)}(\nu,s_0,\mu_R)}{F_L(\nu)}=\frac{\chi(\nu)}{2}\ln\frac{s_0}{Q^2}
+\frac{\beta_0}{4 N_c}\ln\frac{\mu_R^2}{Q^2}
+\frac{3 C_F}{4 N_c}-\frac{5}{18}\frac{n_f}{N_c}+\frac{\pi^2}{4}+\frac{85}{36}
-\frac{\pi^2}{\cosh^2(\pi\nu)}
\]
\beq
\label{ifL}
-\frac{8\left(1+4 i\nu \right)}{\left( 1+2 i\nu \right)^2\left(1-2 i\nu \right)
\left(3 +2 i\nu \right)}+\frac{4}{3-4 i\nu +4\nu^2}\chi\left(\nu\right)
\eeq
\[
+\frac{1}{2}\chi\left(\nu\right)\left[\psi\left(\frac{1}{2}-i\nu\right)
+2\psi\left(\frac{3}{2}-i\nu\right)-2\psi\left(3-2 i\nu\right)
-\psi\left(\frac{5}{2}+i\nu\right)\right]\;.
\]

\section{Numerical analysis and discussion}

We compared several different representations
of the NLA $\gamma^* \gamma^*$ total cross section, differing one from
the other by terms beyond the NLA, and confront them with 
the experimental data from the OPAL and L3 experiments at LEP2, taking
equal photon virtualities, $Q_1=Q_2\equiv Q$, with $Q^2$=17 GeV$^2$, and 
energy range $Y=2\div 6$, where $Y\equiv\ln(s/Q^2)$.
We considered the following representations: (i) Chirilli-Kovchegov 
representation, based on the expression for the cross section as given in 
Ref.~\cite{Chirilli2014}, calculated at the ``natural'' scales $s_0=\mu_R^2=Q^2$;
(ii) series representation with PMS optimization; (iii) exponential 
representation with PMS optimization; (iv) exponential representation with BLM 
optimization, in the two variants $(a)$ and $(b)$ discussed in 
Ref.~\cite{us_BLM}.

\begin{figure}[tb]
\centering
\includegraphics[scale=0.65]{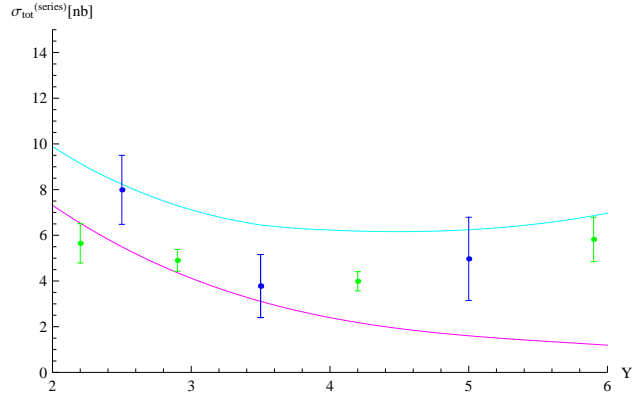}
\caption{$\sigma^{\rm (series)}_{\rm tot}$ {\it versus} $Y$ at
$Q^2=17$ GeV$^2$ ($n_f=4$) (magenta line), together with the experimental
data from OPAL (blue points, $Q^2=18$ GeV$^2$) and L3 (green points,
$Q^2=16$ GeV$^2$); the cyan line represents the result of
Ref.~\cite{Caporale2008} (see Fig.~3 there).}
\label{seriesPMS}
\end{figure}

In all cases, the quark box contribution was included. For the relevant
formulas, tables, plots and other details, we refer to~\cite{IMP14}. 
We show in Fig.~\ref{seriesPMS} the case of the series representation
with PMS optimization, which is representative of all other cases: the effect 
of the BFKL resummation is small and changes only by little the determination 
coming from the LO quark box diagrams. This means that, in the considered range 
of energies, the NLO corrections to the photon IF compensate almost exactly 
the LO ones. Indeed, previous estimates of the cross 
section~\cite{Brodsky,Caporale2008,Zheng} 
using LO IFs together with the NLA BFKL Green's function showed a 
better agreement with LEP2 data.

Reasons for the disagreement in the range $Y=3.5\div 6$ could be:
(i) the BFKL contribution still does not dominate over terms which are 
suppressed by powers of the energy $\sim 1/s$, not included in the present 
consideration; (ii) the presumably large effects in the next-to-NLA are not 
under satisfactory control by the representations of the cross section and 
by the optimization methods we considered. In both cases, the source 
of the problems is in the large negative value of NLO contributions to the 
photon IF. Indeed, in the region of $\nu\simeq 0$, which dominates the
$\nu$-integral in the cross section, the ratio NLO/LO for 
the photon IF is more negative than for the photon-to-meson IF
(see Ref.~\cite{IMP14}).

Another issue is the following: the NLO photon IFs as extracted 
from~\cite{Chirilli2014} have very simple subleading $\sim 1/N_c^2$ 
contributions, in sharp contrast with the cases of NLO photon-to-meson 
IF~\cite{IKP04} and NLO forward jet IF~\cite{MN_IF}. It would be interesting to 
understand the reason for this practically complete cancellation of the 
subleading $1/N_c^2$ terms.
Finally, the photon IF used in this paper (derived from the results 
in~\cite{Balitsky2012,Chirilli2014}) and the one obtained in the conventional 
BFKL approach by Bartels and collaborators~\cite{gammaIF}, presented
in~\cite{Chachamis:2006zz} for the case of transverse polarization, have 
a very different behavior in the variable $x$ (the dimensionless 
ratio of the Reggeon transverse momentum and the photon virtuality squared); 
a qualitative agreement could be obtained reducing the NLO result given in 
Eq. ~(\ref{ifT}) by the factor $\sim 1.87$. It is important and urgent that 
the authors of~\cite{Chachamis:2006zz} finally publish their results for the 
photon IF, since it would be an independent test of the results 
obtained by Balitsky and Chirilli in a completely different approach.

\begin{theacknowledgments}
The work of D.I. was  supported in part by the grant RFBR-13-02-00695-a.
The work of B.M. was supported by the European Commission, European Social
Fund and Calabria Region, that disclaim any liability for the use that can be
done of the information provided in this paper.
\end{theacknowledgments}

\bibliographystyle{aipproc}   

\end{document}